\pdfoutput=1
\documentclass[iop,twocolumn,apjl,letterpaper]{emulateapj}

\usepackage{graphicx}
\usepackage{amsmath}
\newcommand       \rhocgs       {{\rm g~cm}^{-3}}

\begin{document}
\title{To Cool is to Accrete:\\ Analytic Scalings for Nebular Accretion of Planetary Atmospheres}

\author{Eve J. Lee\altaffilmark{1}, Eugene Chiang\altaffilmark{1,2}}
\altaffiltext{1}{Department of Astronomy, University of California Berkeley, Berkeley, CA 94720-3411, USA; evelee@berkeley.edu, echiang@astro.berkeley.edu}
\altaffiltext{2}{Department of Earth and Planetary Science, University of California Berkeley, Berkeley, CA 94720-4767, USA}

\begin{abstract}
Planets acquire atmospheres
from their parent circumstellar disks.
We derive a general analytic expression
for how the atmospheric mass grows with
time $t$, as a function of the underlying core mass
$M_{\rm core}$ and nebular conditions, including
the gas metallicity $Z$.
Planets accrete as much gas as
can cool:
an atmosphere's doubling time is given by its
Kelvin-Helmholtz time. 
Dusty atmospheres behave differently
from atmospheres made dust-free by grain
growth and sedimentation.
The gas-to-core mass ratio (GCR)
of a dusty atmosphere scales as 
GCR~$\propto t^{0.4} M_{\rm core}^{1.7}
Z^{-0.4} \mu_{\rm rcb}^{3.4}$, where
$\mu_{\rm rcb} \propto 1/(1-Z)$ (for $Z$ not
too close to 1) is the mean molecular weight
at the innermost radiative-convective
boundary. This scaling applies 
across all orbital distances and
nebular conditions for dusty atmospheres;
their radiative-convective boundaries, which regulate cooling,
are not set by the external environment, but rather by
the internal microphysics of dust sublimation, 
H$_2$ dissociation, and the formation of H$^-$.
By contrast, dust-free
atmospheres have their radiative boundaries
at temperatures $T_{\rm rcb}$
close to nebular temperatures
$T_{\rm out}$, and grow faster at larger orbital
distances where cooler temperatures, and by
extension lower opacities, prevail.
At 0.1 AU in a gas-poor nebula,
GCR~$\propto t^{0.4}
T_{\rm rcb}^{-1.9} M_{\rm core}^{1.6}
Z^{-0.4} \mu_{\rm rcb}^{3.3}$, while
beyond 1 AU in a gas-rich nebula, 
GCR~$\propto t^{0.4}
T_{\rm rcb}^{-1.5} M_{\rm core}^1
Z^{-0.4}\mu_{\rm rcb}^{2.2}$. 
We confirm
our analytic scalings against
detailed numerical models for objects
ranging in mass from Mars ($0.1 M_\oplus$) to
the most extreme super-Earths (10--$20 M_\oplus$),
and explain why heating from
planetesimal accretion cannot 
prevent the latter from undergoing runaway
gas accretion. 
\end{abstract}

\section{Introduction}
\label{sec:intro}

The {\it Kepler} mission
has discovered that at least $\sim$50\%
of Sun-like stars 
harbor ``super-Earths"---here
defined as planets having radii
1--$4R_\oplus$ \citep[e.g.,][]{fressin13}.\footnote{What we call
``super-Earths" are sometimes
sub-divided into ``super-Earths" and 
the larger ``mini-Neptunes". We do not make
this distinction here.}
Their masses, measured by transit
timing variations \citep[e.g.,][]{hadden14}
and Doppler radial velocities \citep[e.g.,][]{weiss14}, 
imply bulk densities
that are typically $\lesssim 3$ g cm$^{-3}$.
These densities are too low to be compatible with
a pure rock composition. The consensus
view \citep[see also][]{rogers15}
is that the masses of super-Earths
are dominated by their solid cores ---
of mass $M_{\rm core} \simeq 2$--$20 M_\oplus$
and radius $R_{\rm core} \simeq 1$--$2 R_\oplus$ --- while their total radii can be more
than doubled by voluminous, hydrogen-rich atmospheres.
Interior models suggest gas-to-core mass
ratios (GCRs) up to $\sim$10\% \citep[e.g.,][]{lopez14}
and more typically
$\sim$1\% \citep{wolfgang15}.

Unlike the Earth's atmosphere (GCR $\sim 10^{-6}$),
the atmospheres of extrasolar super-Earths are
likely too massive to have been outgassed from
rock~\citep[e.g.,][]{rogers10}. More plausibly,
super-Earth atmospheres originated
as the envelopes of gas giants like Jupiter
did, by accretion from the primordial nebula.
Studies of nebular accretion
\citep[e.g.,][]{ikoma12,bodenheimer14,inamdar15} 
find that super-Earth cores can accrete 
atmospheres having GCRs of $\sim$1--10\% 
before the gas disk dissipates on Myr timescales.
Even higher GCRs can be obtained under
a variety of
conditions \citep{paper1}.
These higher values may be required because
once the parent nebula disperses
and planets are laid bare,
atmospheric loss driven by stellar
irradiation \citep[e.g.,][]{lopez13,owen13}
and by the young planet's heat of formation
\citep{owen15}
can pare GCRs down
by factors of several or more.

\citet{paper1}, hereafter paper I, 
computed nebular accretion histories for a range of
core masses, disk temperatures and densities,
and atmospheric metallicities and dust contents.
Our aim here, in paper II, is to 
provide an analytic understanding of their
numerical results.
Benefiting from hindsight,
we will reduce our model to a few
essential elements and obtain simple
power-law scalings between GCR, time $t$,
core mass $M_{\rm core}$, and metallicity $Z$.
These scalings will be derived against
a variety of backdrops: 
gas-rich vs.~gas-poor nebulae;
dusty vs.~dust-free atmospheres;
close-in vs.~far-out orbital
distances.

Before we present these scaling relations
(Section \ref{sec:GCR}), we revisit the 
fundamental assumption
underlying them: that the nascent
atmospheres have no power source
but passively cool by radiating into 
their nebular wombs.
One source of power that we ignore ---
but that is commonly invoked
in the literature --- is
the accretion of planetesimals.
Paper I provided reasons why planetesimal
accretion could plausibly be dropped
when considering the origin of super-Earth
atmospheres; in Section \ref{ssec:Lacc}
below, we flesh these arguments out more fully
and quantitatively. Readers interested
in the main results of this paper can skip
ahead to Section \ref{sec:GCR}
which derives the GCR $(t, M_{\rm core}, Z)$
scalings for passively cooling atmospheres,
and to Section \ref{sec:conclusion}
which contains a recapitulation with commentary.

\subsection{Planetesimal Accretion}
\label{ssec:Lacc}

Paper I articulated one of the main puzzles posed
by super-Earths: how, despite their large
core masses, they avoided being transformed
into Jupiter-mass giants by runaway gas accretion,
and instead had their GCRs
stabilized at values of several percent.

One way to stop an atmosphere from growing
is to supply it with sufficient
heat --- enough to balance cooling and arrest
secular contraction.
The energy released by the accretion of
planetesimals is a candidate heat source.
We can estimate the required
rates of mass delivery $\dot{M}$
by equating the accretion luminosity
\begin{equation} \label{eq:lacc}
L_{\rm acc} = \frac{GM_{\rm core}\dot{M}}{R_{\rm core}}
\end{equation}
to the cooling luminosity $L_{\rm cool}$,
where the latter is computed from the models
presented in paper I. Here $G$ is the gravitational
constant. Equation (\ref{eq:lacc})
assumes that planetesimals are large enough to
penetrate the atmosphere and release their kinetic
energy at the core surface.

\begin{figure}[!tbh]
\centering
\includegraphics[width=0.4\textwidth]{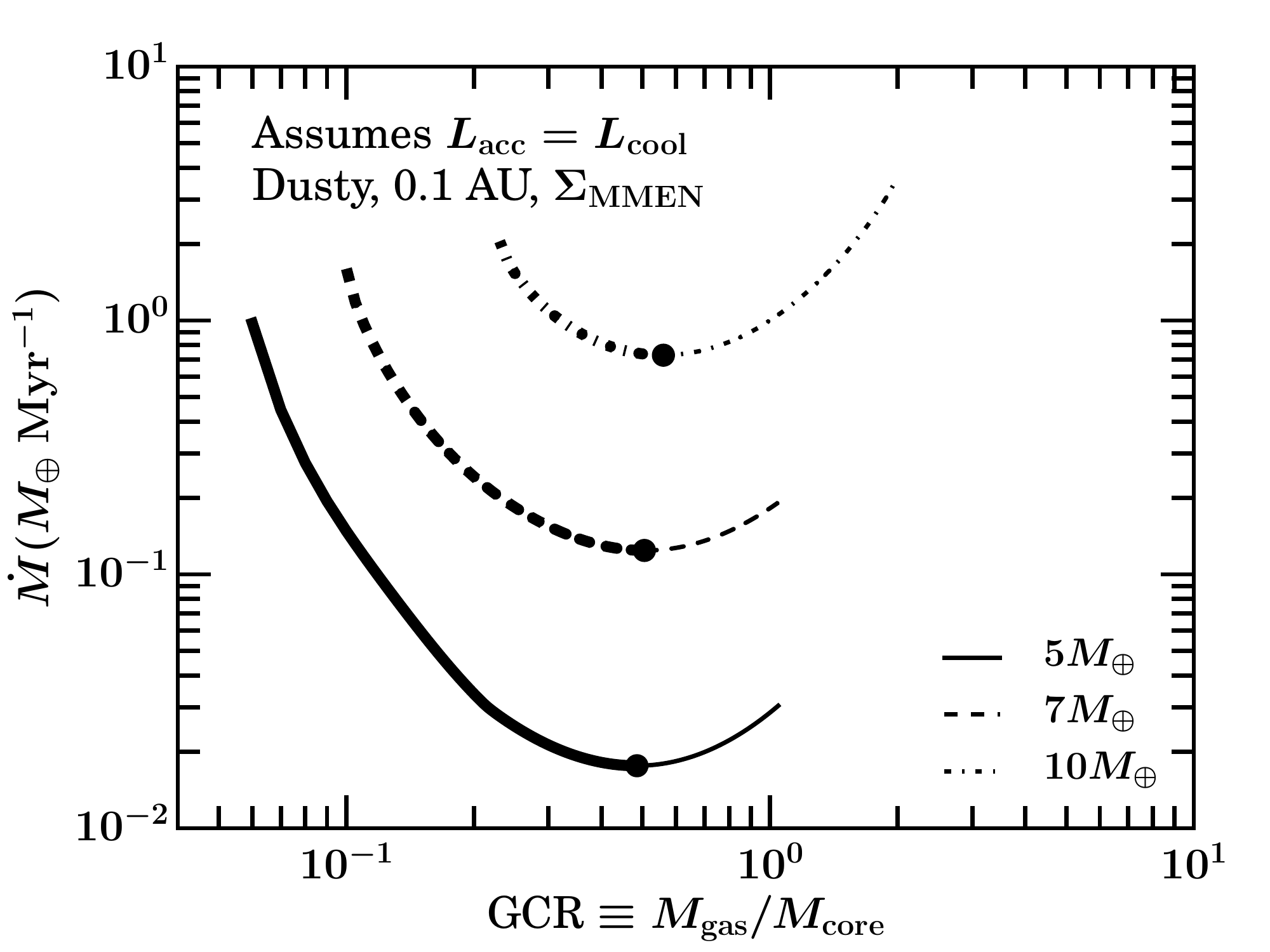}
\caption{Planetesimal accretion rates required
to keep dusty 
envelopes in thermal
equilibrium at a given gas-to-core mass
ratio GCR.
Mass infall rates $\dot{M}$ are calculated 
by equating the accretion luminosity $L_{\rm acc}$
to the cooling luminosity $L_{\rm cool}$,
where the latter is computed from our numerical
models of dusty, passively cooling atmospheres in a
minimum-mass extrasolar nebula (MMEN)
at 0.1 AU (paper I; for reference,
the gas surface density $\Sigma$
of the MMEN is several times larger than
that of the traditional minimum-mass solar nebula;
the difference is immaterial for all the results
presented in this paper).
Circles mark minima in $\dot{M}$ which
correspond to minima in $L_{\rm cool}$;
at these minima, GCR $\sim 0.5$ and envelopes
are on the brink of runaway gas accretion.
To the left of these minima, at smaller GCRs, are 
stable equilibria (thick lines) and to the right 
are unstable equilibria (thin lines).
To stabilize GCRs at values $\lesssim 0.5$
requires that planetesimal accretion rates
be fine-tuned to the values plotted.
In the case of cores of mass $10 M_\oplus$,
this fine-tuning would still be unable to
prevent runaway: the required
$\dot{M}$'s, of order $\sim$1 $M_\oplus$
Myr$^{-1}$, would double the core mass within
a disk lifetime of $\sim$10 Myr and push
the cores over to runaway (see also Figure 
\ref{fig:Lratio_Mdot}).
}
\label{fig:Mdot_v_gcr_full}
\end{figure}

Figure \ref{fig:Mdot_v_gcr_full}
displays the planetesimal accretion rates $\dot{M}$
so estimated, using the dusty atmosphere
models from paper I (dust-free atmospheres
will be considered shortly).
Every curve exhibits
a minimum in $\dot{M}$ with GCR; this minimum
corresponds to the minimum in
$L_{\rm cool}$ with time that 
appears in all 
passively cooling and growing
atmospheres \citep[see also][]{piso14}
and that we used in paper I
to mark the onset of runaway gas accretion. 
To the left of the minima, 
the increasing atmospheric density with increasing
GCR renders the envelope more opaque and causes
$L_{\rm cool}$ to drop
(see Section 3.1 of paper I and Section
\ref{sec:GCR} of this paper).
To the right of the minima, at GCR $\gtrsim 0.5$,
the self-gravity of 
the gas envelope becomes significant, 
and larger $L_{\rm cool}$ is required to 
balance stronger gravity.

The curves in Figure \ref{fig:Mdot_v_gcr_full} 
represent the loci of thermal equilibrium: 
atmospheres are stabilized at a given GCR 
when solids rain down at the corresponding $\dot{M}$.
We now consider the stability of these equilibria.
To the left of the minima, at lower GCR, 
equilibria are stable; for example,
a perturbation to higher GCR decreases
$L_{\rm cool}$ below $L_{\rm acc}$ (which is presumed fixed),
resulting in a net heating that expands the atmosphere
and lowers the GCR back down to its equilibrium value.
To the right of the minima, at higher GCR, there are no
stable equilibria because
$L_{\rm cool}$ increases with both GCR
and $M_{\rm core}$.
As $M_{\rm core}$ increases
from planetesimal accretion,
$L_{\rm cool}$ rises rapidly (see Section 3.2.3 of
paper I and Section \ref{sec:GCR} of this paper),
outpacing $L_{\rm acc}$ and triggering runaway.

The values of $\dot{M}$ to the left
of the minima in Figure \ref{fig:Mdot_v_gcr_full}
represent possible solutions to the puzzle
of how super-Earths avoided runaway gas
accretion. But we find these solutions
unsatisfactory for a couple of reasons.
The first is that planetesimal accretion rates
must be fine-tuned to the values
plotted. Why, for example, should $5M_\oplus$ cores
accrete $0.1 M_\oplus$/Myr in 
solids to have their GCRs stabilized at $\sim$0.1?
Planetesimal accretion rates
are influenced by a host of factors
\citep[e.g.,][]{goldreich04},
and most estimates 
lead to rates orders of magnitude higher
than the ones plotted
in Figure \ref{fig:Mdot_v_gcr_full}
\citep[see, e.g., Appendix A of][]{rafikov06}.
The much higher accretion rates are natural
consequences of
the high disk surface densities and short
orbital times characterizing the small
orbital distances where {\it Kepler}
super-Earths are found, and lead
us to surmise that planetesimals are
fully incorporated into planets on timescales
much shorter than the 1--10 Myrs
required for atmospheres to cool
and grow.\footnote{Our discussion
here pertains to ``planetesimals", not
to the larger protocores which are thought
to assemble into super-Earth cores by
``major mergers". Such mergers take place over
a wide range of timescales that can overlap
or even exceed gas disk lifetimes, depending
sensitively on the disk's solid surface density
\citep{dawson15}. Energy release from
the last major merger could, in principle, provide
a significant source of heat to stop
gas accretion \citep{inamdar15}, 
but only if the core transports its internal
energy outward on a timescale comparable to the
atmospheric cooling time of 1--10 Myr.
The actual energy transport timescale of the core 
is highly uncertain, depending on the 
unknown viscosity (see paper I, section 3.1.2).}

\begin{figure}[!tbh]
\centering
\includegraphics[width=0.4\textwidth]{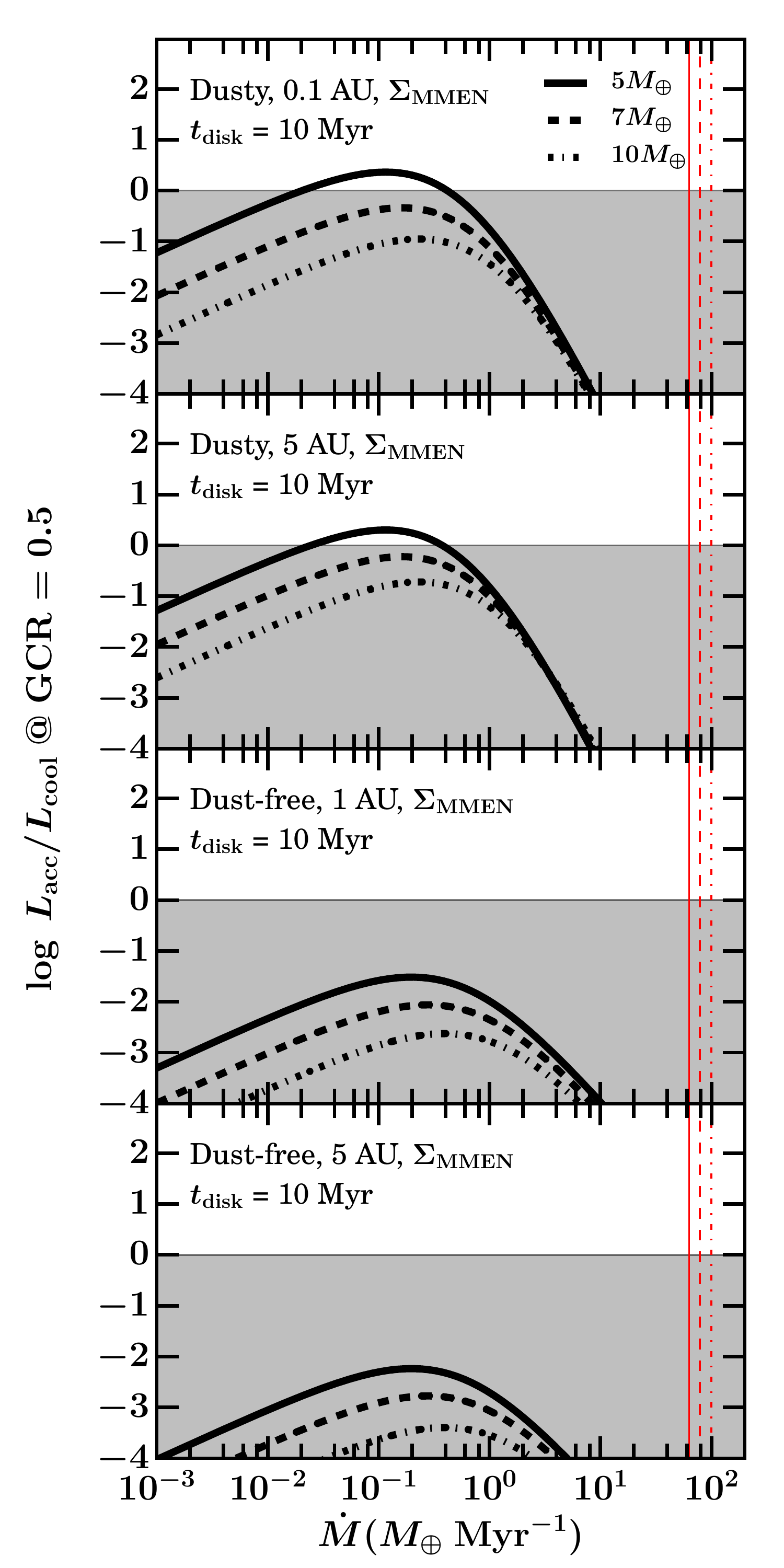}
\caption{Preventing runaway with heating from
infalling planetesimals is possible only for 
specific ranges of mass infall rates $\dot{M}$
for sufficiently low-mass cores.
Accounting for the growth of the core mass from
planetesimal accretion (curves
are labeled by initial core masses),
we plot the ratio of $L_{\rm acc}$ to
$L_{\rm cool}({\rm GCR}=0.5)$ 
after a gas lifetime of $t_{\rm disk} = 10$ Myr
in a gas-rich (MMEN) disk,
considering both dusty and dust-free 
atmospheres from 0.1 to 5 AU (see annotation in each panel).
If $L_{\rm acc}/L_{\rm cool} \geq 1$ (unshaded),
then runaway is successfully avoided (see equation
\ref{eq:Lacc} and surrounding discussion).
All four panels show that the ranges 
of $\dot{M}$ and initial core mass
required to avoid gas giant formation are extremely
limited --- if they exist at all. 
If $\dot{M}$ is too low, 
$L_{\rm acc}$ can never balance $L_{\rm cool}$, 
and if $\dot{M}$ is too high, cores rapidly grow in
mass and become giants within the
disk lifetime. 
The lowest $\dot{M}$'s estimated
from first principles by 
\citet[][see his equation A1]{rafikov06} are indicated by red lines; these
$\dot{M}$'s
readily push cores over to runaway.
}
\label{fig:Lratio_Mdot}
\end{figure}

The second reason we do not find planetesimal
accretion attractive is that
it bites off more than it can chew:
Figure \ref{fig:Mdot_v_gcr_full} implies
that a $10 M_\oplus$ core has to accrete
planetesimals at such a high rate that it
doubles in mass within the disk
lifetime of $t_{\rm disk} \sim 10$ Myr.
The $20M_\oplus$ core that results is
certain to undergo
runaway gas accretion
in a disk that is gas-rich (i.e.,
with a gas surface density $\Sigma$ comparable to that
of the minimum-mass solar nebula).

We can illustrate with another calculation
how planetesimal accretion leads to problems
of fine-tuning and runaway gas accretion. 
An atmosphere does not grow if
$L_{\rm acc} > L_{\rm cool}$.
In particular, runaway is avoided
if $L_{\rm acc} > L_{\rm cool}$ 
when ${\rm GCR} \simeq 0.5$, 
the critical GCR above which 
envelope self-gravity starts becoming significant.
We therefore
evaluate the ratio $L_{\rm acc}/L_{\rm cool}$
when GCR = 0.5, presuming the core has
accreted planetesimals at a constant
rate for the preceding time $t_{\rm disk}$:
\begin{equation}
\label{eq:Lacc}
\left.\frac{L_{\rm acc}}{L_{\rm cool}} \right|_{{\rm GCR}=0.5} = \frac{G(M_{\rm core} + \dot{M}t_{\rm disk})\dot{M}}{R_{\rm core}L_{\rm cool}}
\end{equation}
where $L_{\rm cool}$ is evaluated using
our atmospheric models for the final core
mass $M_{\rm core} + \dot{M}t_{\rm disk}$.
(This evaluation uses the scaling relation
between $L_{\rm cool}$ and $M_{\rm core}$
as derived in Section \ref{sec:GCR} below.)
Figure \ref{fig:Lratio_Mdot} plots
$L_{\rm acc}/L_{\rm cool}|_{{\rm GCR}=0.5}$
against $\dot{M}$.
Cores of initial mass $M_{\rm core} = 5 M_\oplus$
can avoid runaway --- but only if 
their atmospheres are dusty, and if $\dot{M}$
is tuned to a narrow range.
This window closes completely for
cores of initial mass $\gtrsim 7M_\oplus$:
such cores will inevitably undergo runaway within
10 Myr in gas-rich disks,
regardless of the magnitude of planetesimal
accretion or their proximity to their central stars.
These conclusions are only amplified for
cores with dust-free atmospheres.
We will elaborate in Section
\ref{sec:GCR} on
these trends with nebular environment
and atmospheric composition.
The point here is that
planetesimal accretion does not generically
prevent runaway. 
Avoiding runaway for even low-mass cores
with $M_{\rm core} < 5 M_\oplus$
requires delicate adjustment of $\dot{M}$
which seems difficult
to achieve
under general circumstances.
We are motivated therefore to study
gas accretion histories that omit
planetesimal accretion --- this is the
main subject of this paper, to which we
now turn.

\section{Scaling Relations for GCR ($M_{\lowercase{\rm core}},\lowercase{t},Z)$}
\label{sec:GCR}

We derive general scaling relations
for how the gas-to-core mass ratio GCR
varies with core mass $M_{\rm core}$,
time $t$, and metallicity $Z$, 
in the absence of external heating.
The derivation is semi-analytic in that a
few parameters will be calibrated using our
numerical models (see paper I for details 
on how we build our numerical models).

Accretion is mediated by cooling: upon
radiating away its energy,
a planet's atmosphere contracts, allowing
nebular gas to refill the Hill sphere.
The system self-regulates so that whatever
atmosphere of mass $M_{\rm gas}$
has been accreted has a cooling
time equal to the time that has elapsed:
\begin{equation}
\frac{M_{\rm gas}}{\dot{M}_{\rm gas}} \sim t \sim t_{\rm cool} \sim \frac{|E|}{L_{\rm cool}}
\label{eq:t}
\end{equation}
where $E$ is the atmosphere's total energy
and $L_{\rm cool}$ is its luminosity. Statement
(\ref{eq:t}) is perhaps more easily
understood by considering
the inverse cases 
$t \ll t_{\rm cool}$
(GCR is overestimated because
not enough time has elapsed to accrete
such a thick atmosphere)
and $t \gg t_{\rm cool}$
(GCR is underestimated because there is
plenty of time for the atmosphere to
continue cooling
and growing).

The relevant cooling time is that of the
innermost convective zone which contains
most of the atmosphere's mass and energy.
To estimate $E$, we use the fact that
in hydrostatic equilibrium, an atmospheric
mass $M_{\rm gas} \equiv {\rm GCR} \times
M_{\rm core}$ has a total energy of order
its gravitational potential energy:
\begin{equation}
\label{eq:E1}
|E| \sim \frac{GM_{\rm core}M_{\rm gas}}{R}
\end{equation}
where $G$ is the gravitational constant.
What $R$ should we choose: the core radius
$R_{\rm core}$ 
or the radiative-convective boundary $R_{\rm rcb}$?\footnote{The outer radius
of our numerical models is either the Bondi or Hill
radius, whichever is smaller. Neither of these
radii enters into our analytic theory,
since not much mass is situated near the outer boundary.}
The answer depends on how steep the density
profile is.
For fixed adiabatic index $\gamma$,
the density profile in the isentropic
convective zone follows
\begin{equation}
\label{eq:rho1}
\rho = \rho_{\rm rcb}\left[1+\nabla_{\rm ad}\frac{GM_{\rm core}}{c^2_{\rm rcb}}\left(\frac{1}{r} - \frac{1}{R_{\rm rcb}}\right)\right]^{1/(\gamma-1)}
\end{equation}
where $\rho_{\rm rcb}$ is the density at the
radiative-convective boundary (rcb), 
$\nabla_{\rm ad} = (\gamma-1)/\gamma$ is the
adiabatic gradient, $c_{\rm rcb}^2 \equiv 
kT_{\rm rcb}/\mu_{\rm rcb} m_{\rm H}$,
$T_{\rm rcb}$ and $\mu_{\rm rcb}$ are the
temperature and mean molecular weight
evaluated at the rcb,
$k$ is Boltzmann's constant, and $m_{\rm H}$
is the atomic mass of hydrogen.
Now because
$1/r > 1/R_{\rm rcb}$ and $GM_{\rm core}/c_{\rm rcb}^2r 
> GM_{\rm core}/c_{\rm rcb}^2R_{\rm rcb} \sim 1$ 
(the last equality follows from hydrostatic 
equilibrium), equation (\ref{eq:rho1}) can be approximated as
\begin{equation}
\label{eq:rho}
\rho \sim \rho_{\rm rcb}\left(\nabla_{\rm ad}\frac{R_{\rm b,rcb}}{r}\right)^{1/(\gamma-1)}
\end{equation}
where $R_{\rm b,rcb} \equiv GM_{\rm core}/c_{\rm rcb}^2$. Equation (\ref{eq:rho}) implies that
if $\gamma < 4/3$, then 
the atmosphere's mass is concentrated near
$R_{\rm core}$ rather than near $R_{\rm rcb}$.
We find that the convective zones of all our numerical models
are indeed
characterized by $\gamma \leq 4/3$: the adiabatic
gradient drops at temperatures exceeding 2500 K
as energy is spent dissociating H$_2$ 
rather than heating the gas.
Therefore we choose $R = R_{\rm core}$ in equation 
(\ref{eq:E1}):\footnote{This is contrary to equation 
(32) in paper I which mistakenly
assumes the atmosphere's mass 
is concentrated near $R_{\rm rcb}$ instead of near
$R_{\rm core}$. 
The correction lengthens the runaway time estimated in that equation
by a factor of 10, bringing it into
closer agreement with the numerical result cited there.}
\begin{align}
\label{eq:Efinal}
|E| &\sim \frac{GM_{\rm core}^2 \times {\rm GCR}}{R_{\rm core}} \nonumber \\
&\sim G\left(\frac{4\pi\rho_{\rm b}}{3}\right)^{1/3}M_{\rm core}^{5/3} \times {\rm GCR} \nonumber \\
&\sim f_E M_{\rm core}^{5/3} \times {\rm GCR} 
\end{align}
where $f_E \equiv G(4\pi\rho_{\rm b}/3)^{1/3}$ 
and $\rho_{\rm b}$
is the bulk density of the core (assumed constant for this paper;
we neglect the small variation of core density
with core mass; see, e.g.,
\citealt{valencia06} and \citealt{fortney07}).

We now examine $L_{\rm cool}$. 
The rcb controls the rate at which 
the innermost convective zone cools. 
Very little luminosity is generated above the rcb
(as verified in paper I, section 3.1.2),
so we evaluate $L_{\rm cool}$ at the rcb:
\begin{equation}
\label{eq:L1}
L_{\rm cool} = \frac{64 \pi G (1+ {\rm GCR})M_{\rm core}\sigma T_{\rm rcb}^3 \mu_{\rm rcb} m_{\rm H} \nabla_{\rm ad}}{3k\rho_{\rm rcb}\kappa_{\rm rcb}}
\end{equation}
where $\sigma$ is the Stefan-Boltzmann
constant 
and $\kappa_{\rm rcb}$ is the opacity at the rcb.
We parameterize the latter as
\begin{equation} \label{eq:kappapar}
\kappa_{\rm rcb} = 
\kappa_0 (\rho_{\rm rcb}/\rho_0)^{\alpha} 
(T_{\rm rcb}/T_0)^{\beta} (Z/Z_0)^{\delta}
\end{equation}
where the various
constants depend on microphysics which vary from case to case (details to be given in the
subsections below).

We relate $\rho_{\rm rcb}$ to GCR as follows.
The total atmospheric mass in the inner convective zone is
\begin{align}
\label{eq:Matm1}
M_{\rm gas} &= 4\pi\int_{R_{\rm core}}^{R_{\rm rcb}}r^2\rho(r)dr \nonumber \\
            &\sim 4\pi\rho_{\rm rcb}(\nabla_{\rm ad}R_{\rm b,rcb})^{1/(\gamma-1)}R_{\rm core}^{3-1/(\gamma-1)}
\end{align}
where we substituted (\ref{eq:rho}). Then
\begin{equation}
\label{eq:rhorcb}
\rho_{\rm rcb} \sim \frac{{\rm GCR} \times M_{\rm core}}{4\pi(\nabla_{\rm ad}R_{\rm b,rcb})^{1/(\gamma-1)}R_{\rm core}^{3-1/(\gamma-1)}}.
\end{equation}

Substituting (\ref{eq:kappapar}) and
(\ref{eq:rhorcb}) into (\ref{eq:L1}):
\begin{align}
\label{eq:L}
L_{\rm cool} &\sim \frac{64\pi G \sigma m_{\rm H}}{3k\kappa_0\rho_0^{-\alpha} T_0^{-\beta}} \left(\frac{Z_0}{Z}\right)^{\delta} \frac{T_{\rm rcb}^{3-\beta}\mu_{\rm rcb} \nabla_{\rm ad} (1+{\rm GCR})M_{\rm core}}{\rho_{\rm rcb}^{1+\alpha}} \nonumber \\
&\sim \frac{4^{4+\alpha}\pi^{2+\alpha}G\sigma m_{\rm H}}{3k\kappa_0\rho_0^{-\alpha} T_0^{-\beta}}\left(\frac{Z_0}{Z}\right)^{\delta} \frac{T_{\rm rcb}^{3-\beta}\mu_{\rm rcb} \nabla_{\rm ad} (1+{\rm GCR})}{{\rm GCR}^{1+\alpha}} \nonumber \\
&\,\,\,\,\,\,\,\,\times M_{\rm core}^{-\alpha}(\nabla_{\rm ad} R_{\rm b,rcb})^{\frac{1+\alpha}{\gamma-1}} R_{\rm core}^{(3-\frac{1}{\gamma-1})(1+\alpha)} \,.
\end{align}
We re-write (\ref{eq:L}) in terms of $M_{\rm core}, Z, \mu_{\rm rcb}$, and GCR: 
\begin{align}
\label{eq:Lfinal}
L_{\rm cool} &\sim f_L \left(\frac{Z_0}{Z}\right)^{\delta} T_{\rm rcb}^{3-\beta-\frac{1+\alpha}{\gamma-1}}\frac{1+{\rm GCR}}{{\rm GCR}^{1+\alpha}} \nonumber \\
& \,\,\,\,\,\,\,\,
\times (\mu_{\rm rcb}\nabla_{\rm ad})^{1+\frac{1+\alpha}{\gamma-1}} M_{\rm core}^{1+\frac{2}{3}\left(\frac{1+\alpha}{\gamma-1}\right)}
\end{align}
where 
\begin{align}
f_L &\equiv \frac{4^{4+\alpha}\pi^{2+\alpha}\sigma}{3\kappa_0\rho_0^{-\alpha} T_0^{-\beta}} \nonumber \\
&\,\,\,\,\,\,\,\times \left(\frac{Gm_{\rm H}}{k}\right)^{1+\frac{1+\alpha}{\gamma-1}} \left(\frac{3}{4\pi\rho_{\rm b}}\right)^{(1+\alpha)\left[1-\frac{1}{3(\gamma-1)}\right]}.
\end{align}

When all the hydrogen is molecular,
the mean molecular weight $\mu$ depends on $Z$ as:
\begin{align}
\label{eq:mu}
\mu &\sim \frac{1}{0.5 X + 0.25 Y + 0.06 Z} \\
X &= \frac{1-Z}{1.4} \nonumber \\
Y &= \frac{0.4(1-Z)}{1.4} \nonumber
\end{align}
where $X$ and $Y$ are the hydrogen and 
helium mass fractions, respectively. 
The prefactor of 0.06 for $Z$ corresponds to
the contribution from atomic metals
using the abundances of \citet{grevesse93};
these abundances are the ones adopted
by \citet{ferguson05}, whose opacities
we use.

Collecting 
(\ref{eq:Efinal}) and (\ref{eq:Lfinal}) into (\ref{eq:t}) yields
\begin{align}
\label{eq:tfinal}
t &\sim \frac{f_E}{f_L}\left(\frac{Z}{Z_0}\right)^{\delta}\frac{{\rm GCR}^{2+\alpha}}{1+{\rm GCR}} M_{\rm core}^{\frac{2}{3} \left( 1-\frac{1+\alpha}{\gamma-1} \right)} \nonumber \\
&\,\,\,\,\,\,\,\times T_{\rm rcb}^{-3+\beta+\frac{1+\alpha}{\gamma-1}} (\mu_{\rm rcb} \nabla_{\rm ad})^{-1-\frac{1+\alpha}{\gamma-1}}
\end{align}
which we invert to arrive at our desired
relation for GCR as a function of
$t, M_{\rm core}, Z,$ and $\mu_{\rm rcb}$,
valid for GCR $\lesssim 1$:\footnote{Retaining
the dependence on $\rho_{\rm b}$ and adopting 
$\rho_{\rm b} \propto M_{\rm core}^{1/4}$~\citep{valencia06}
yields GCR~$\propto M_{\rm core}^{[-1-\alpha/4+3(1+\alpha)/4(\gamma-1)]/(2+\alpha)}$.
This correction hardly changes the dependence of GCR
on $M_{\rm core}$; for example, for dusty atmospheres,
GCR~$\propto M_{\rm core}^{1.8}$ instead of $M_{\rm core}^{1.7}$
(see later subsections).}
\begin{align}
\label{eq:gcr}
{\rm GCR} &= f\left[t \frac{f_L}{f_E} \left(\frac{Z_0}{Z}\right)^{\delta} M_{\rm core}^{\frac{2}{3}\left(\frac{1+\alpha}{\gamma-1}-1\right)} \right. \nonumber \\
&\,\,\,\,\,\,\times \left.(\mu_{\rm rcb} \nabla_{\rm ad})^{1+\frac{1+\alpha}{\gamma-1}} T_{\rm rcb}^{3-\beta-\frac{1+\alpha}{\gamma-1}}\right]^{\frac{1}{2+\alpha}} \,.
\end{align}
We have introduced a dimensionless
fudge factor $f$ which we will normalize
against our numerical models.
The parameters that vary
most from one scenario to another are
the opacity constants in
equation (\ref{eq:kappapar}),
and the rcb variables
$T_{\rm rcb}$ and $\nabla_{\rm ad}$.
All these input constants
will be drawn from our numerical solutions.

The following subsections
examine how equation (\ref{eq:gcr})
plays out in various formation
environments. We consider dusty vs.~dust-free
atmospheres in gas-poor vs.~gas-rich nebulae
at small orbital distances and large.
The plausibility of these scenarios
is not assessed; that exercise is deferred
to a later study (paper III).
In ``dusty'' models, we assume the ISM-like grain size 
distribution of \citet{ferguson05}, 
and in dust-free models, we assume all metals 
to be in the gas phase.
By ``gas-rich" we mean a nebula
whose gas surface density $\Sigma$ equals that of
the minimum-mass extrasolar nebula
(MMEN; see equation 12 of paper I;
for comparison,
the Hayashi \citeyear{hayashi81}
nebula is $\sim$7$\times$ less dense),
and by ``gas-poor" we mean a nebula
whose gas content is $200 \times$ smaller
(one whose gas mass equals its solid mass).
Nebular temperatures are taken from equation
(13) of paper I ($T_{\rm out} = \{1000,400,200\}$ K
at orbital distances $a = \{0.1, 1, 5\}$ AU).

\subsection{Dusty Atmospheres}
\label{ssec:dusty-atm}

Dusty atmospheres are high opacity
atmospheres and tend to be convective
in their upper layers.
They cease being dusty
at depths below which temperatures
are high enough for dust sublimation.
The disappearance of grains causes
the opacity $\kappa$ to drop
by two orders of magnitude; the sudden transparency
opens a radiative window at depth.
This radiative zone appears universally
over all core masses 
and orbital distances as long as the
upper layers are dusty.
The base of
the radiative zone --- i.e., the innermost
radiative-convective boundary --- is located
where H$_2$ dissociates and H$^-$ appears
with its strongly temperature-sensitive
opacity:
\begin{align}
\kappa ({\rm H}^-) & \simeq 3\times 10^{-2}~{\rm cm}^2~{\rm g}^{-1} \left(\frac{\rho}{10^{-4}\, \rhocgs}\right)^{0.5} \nonumber \\
& \times \left(\frac{T}{2500 \,{\rm K}}\right)^{7.5} \left(\frac{Z}{0.02}\right)^1. 
\label{eq:hminus}
\end{align}
Equation (\ref{eq:hminus}), obtained 
by numerically fitting the tabulated
opacities of \citet{ferguson05},
defines the
relevant opacity constants when
evaluating equation (\ref{eq:gcr})
for dusty atmospheres.
Also characterizing dusty models
is $T_{\rm rcb} \simeq 2500$ K:
the temperature at which H$_2$ dissociates.

\subsubsection{Dusty and Gas-Poor from 0.1--1 AU}

Substituting $T_{\rm rcb} = 2500$ K and
the parameters from (\ref{eq:hminus}) into (\ref{eq:gcr}),
and further restricting our attention
to gas-poor nebulae for which
the gas surface density
$\Sigma$ is 1/200 that of the 
MMEN, we find
\begin{align}
\label{eq:gcr_dusty}
{\rm GCR} &\simeq 0.06 \left( \frac{f}{1.2} \right) \left(\frac{t}{1 \,{\rm Myr}}\right)^{0.4} \left(\frac{2500\,{\rm K}}{T_{\rm rcb}}\right)^{4.8} \left(\frac{0.02}{Z}\right)^{0.4} \nonumber \\
&\,\,\,\,\,\,\,\,\left(\frac{\nabla_{\rm ad}}{0.17}\right)^{3.4}\left(\frac{\mu_{\rm rcb}}{2.37}\right)^{3.4} \left(\frac{M_{\rm core}}{5M_\oplus}\right)^{1.7} \,.
\end{align}
Here we have fixed
$\rho_{\rm b} = 7\,\rhocgs$ and $\gamma = 1.2$
(cf.~Figure 3 of paper I which
shows that $\gamma$ ranges from 1.2 to 1.3
inside the innermost convective zone;
although that figure pertains to a gas-rich
nebula, similar values of $\gamma$ are obtained
in a gas-poor nebula).
In writing (\ref{eq:gcr_dusty}), the last parameter to be calculated is the
overall normalization $f$; the best
agreement with our numerical models is
obtained for $f$ between $1.2$ ($a = 0.1$ AU) and
$1.3$ ($a = 1$ AU). (We cannot calibrate
$f$ for $a > 1$ AU in gas-poor nebulae
because the relevant densities fall below
those in our opacity tables.)

\begin{figure}[!tbh]
\centering
\includegraphics[width=0.4\textwidth]{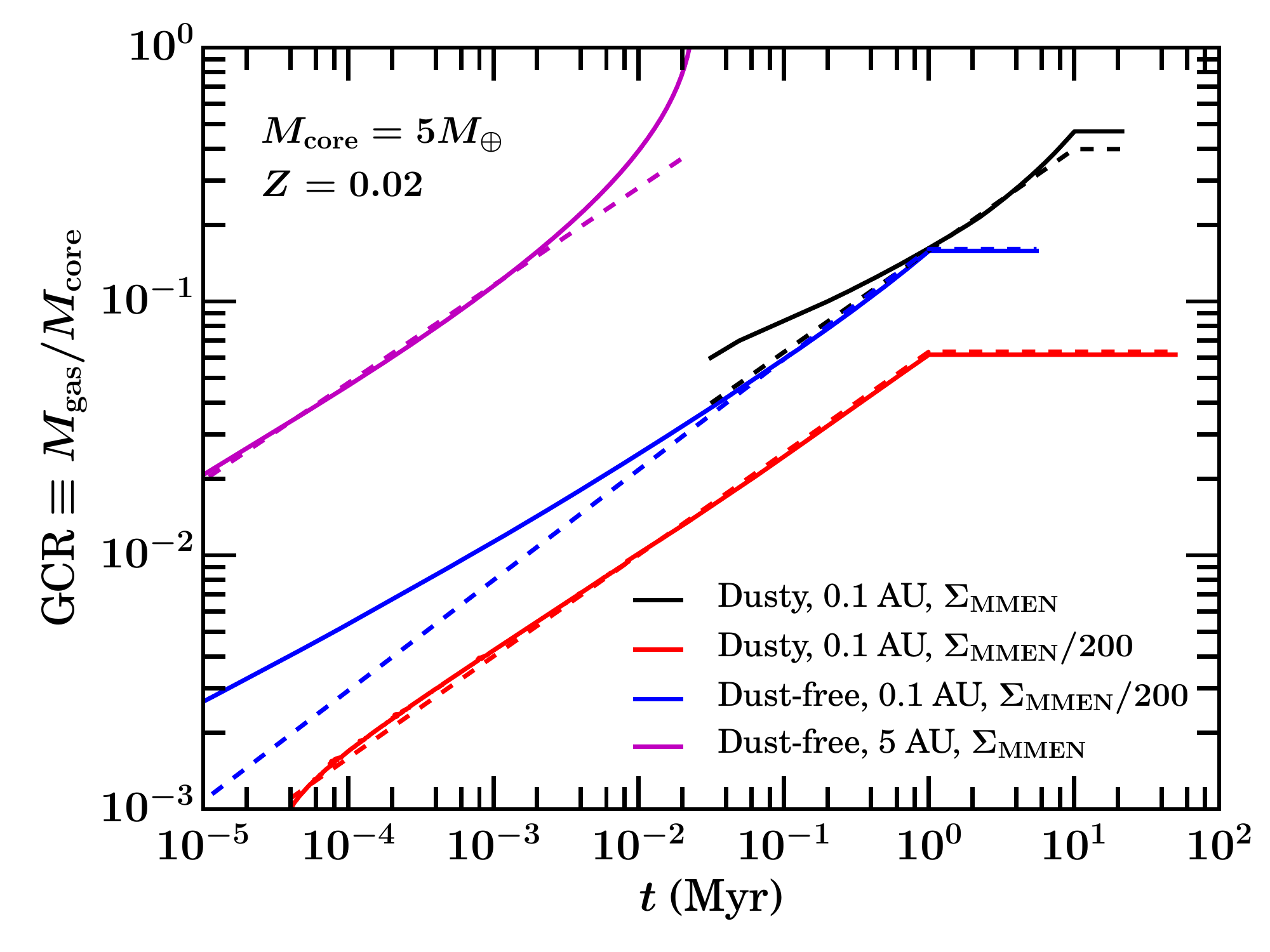}
\caption{Theory (dashed curves;
equations \ref{eq:gcr_dusty}, \ref{eq:gcr_dusty_full},
\ref{eq:gcr-df-0.1au}, and \ref{eq:gcr-df-1au}) vs. numerics (solid
curves) for $5 M_\oplus$
cores under a variety of nebular conditions.
For the most part, the agreement is good:
before runaway,
GCRs do scale with time as $t^{0.4}$ under many
circumstances (see also
the master equation \ref{eq:gcr}).
Solutions are truncated at disk
depletion times: $t_{\rm disk,slow} = 10$ Myr
for gas-rich disks, and $t_{\rm disk,fast} = 1$ Myr
for gas-poor disks. The normalizations for
the semi-analytic curves (i.e., the values of
$f$) are adjusted by hand to match those
of the numerical curves.}
\label{fig:gcr_t}
\end{figure}

\begin{figure}[!tbh]
\centering
\includegraphics[width=0.4\textwidth]{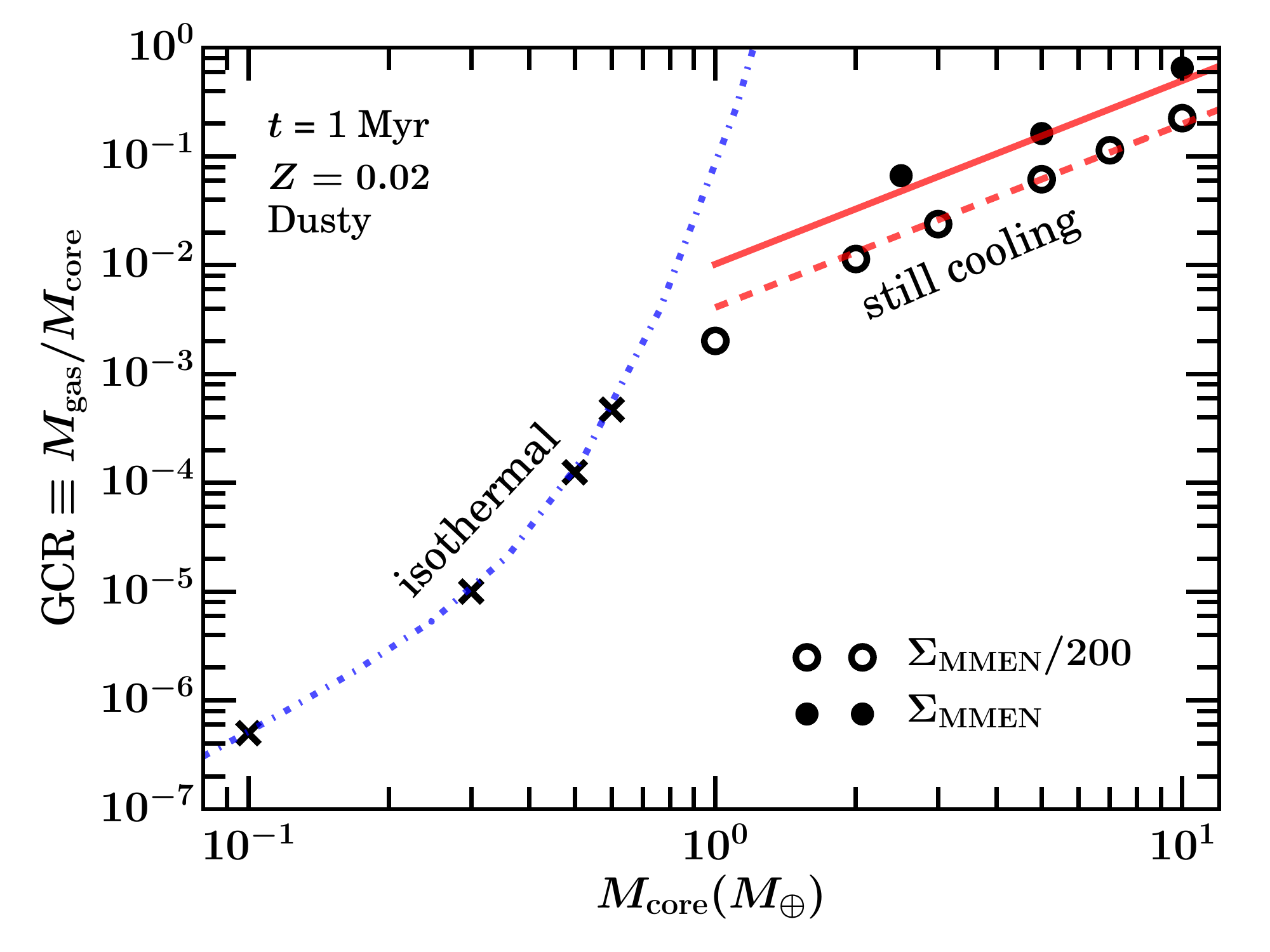}
\caption{\label{fig:gcr_M}
GCR vs.~$M_{\rm core}$ at fixed time $t = 1$ Myr,
demonstrating two stages of atmosphere acquisition.
At low core masses $\lesssim 1 M_\oplus$,
GCRs have reached their maximum values:
atmospheres have ``maximally cooled" to their
isothermal endstates (to cool is to accrete,
and these planets are too cool to accrete further).
Black crosses are numerically calculated
maximum GCRs and match exactly the analytically computed
blue curve for isothermal atmospheres 
(evaluated with $T = 1000$K, the disk temperature at 0.1 AU).
At high core masses $\gtrsim 1 M_\oplus$,
atmospheres are still cooling and growing
at $t = 1$ Myr; their GCRs 
(open and filled circles)
obey the scaling relations
(\ref{eq:gcr_dusty}) and (\ref{eq:gcr_dusty_full})
which predict GCR $\propto M_{\rm core}^{1.7}$.
All data shown are for dusty 
atmospheres at 0.1 AU.}
\end{figure}

\begin{figure}[!tbh]
\centering
\includegraphics[width=0.4\textwidth]{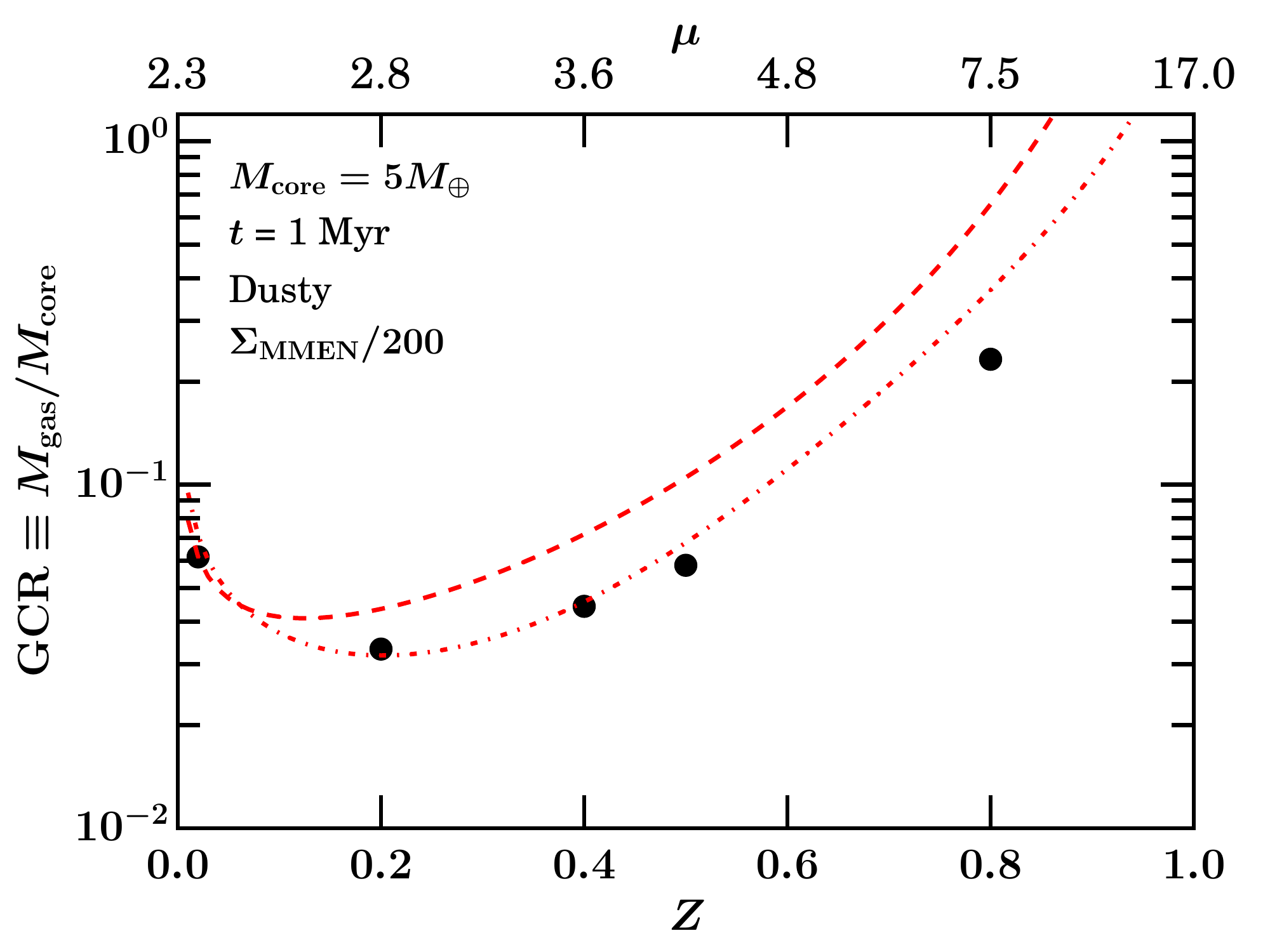}
\caption{GCR vs.~gas metallicity $Z$ at fixed time $t = 1$ Myr
for a $5 M_\oplus$ core in a dusty, gas-poor nebula,
demonstrating that GCR
is not a monotonic function of $Z$
(which itself is assumed constant with time
and space for a given model).
For $Z \lesssim 0.2$, 
the increased opacity with increased $Z$ (equation
\ref{eq:hminus}) suppresses cooling and yields
smaller GCRs.
For $Z \gtrsim 0.2$, increases in mean molecular
weight with $Z$ 
necessitate larger $L_{\rm cool}$ to maintain
hydrostatic equilibrium; faster cooling at higher $Z$
increases GCRs at a given time.
The steep dependence of
${\rm GCR} \propto T_{\rm rcb}^{-4.8}$ predicted by
(\ref{eq:gcr_dusty}) leads us to compute two sets of
curves: one where we fix $\nabla_{\rm ad} = 0.17$ and
$T_{\rm rcb} = 2500$ K (thick dashed),
and another where we let $T_{\rm rcb}$
vary according to the numerical models (thin
dot-dashed) which naturally does better at fitting
the data.}
\label{fig:gcr_Z}
\end{figure}

Figures \ref{fig:gcr_t}, \ref{fig:gcr_M}, and \ref{fig:gcr_Z}
demonstrate how well our semi-analytic scaling 
relation (\ref{eq:gcr_dusty}) does in reproducing
the full numerical results. We emphasize that the 
exponents in equation (\ref{eq:gcr_dusty}) are not 
merely fit parameters, but follow
from the physical considerations
underlying equations (\ref{eq:t})--(\ref{eq:hminus}).

Interestingly, we see in Figure \ref{fig:gcr_M} that 
the dependence of GCR on $M_{\rm core}$ 
at fixed time $t = 1$ Myr differs across $1 M_\oplus$.
Atmospheres atop core masses $\lesssim 1 M_\oplus$
have cooled to their isothermal endstates and
have stopped accreting before the sampled time;
their GCRs have reached their maximum values.

\subsubsection{Dusty and Gas-Rich from 0.1--5 AU}
\label{ssec:gcr-gas-rich}

This case is almost identical to the
case considered above. The only change
in going from gas-poor
($\Sigma=\Sigma_{\rm MMEN}/200$)
to gas-rich
conditions ($\Sigma = \Sigma_{\rm MMEN}$)
is that the fitted normalizations are higher,
running from $f = 3$ (0.1 AU) to 2 (1 AU)
to 1.8 (5 AU):
\begin{align}
\label{eq:gcr_dusty_full}
{\rm GCR} &\simeq 0.16 \left( \frac{f}{3} \right) \left(\frac{t}{1 \,{\rm Myr}}\right)^{0.4} \left(\frac{2500\,{\rm K}}{T_{\rm rcb}}\right)^{4.8} \left(\frac{0.02}{Z}\right)^{0.4} \nonumber \\
&\,\,\,\,\,\,\,\,\left(\frac{\nabla_{\rm ad}}{0.17}\right)^{3.4}\left(\frac{\mu_{\rm rcb}}{2.37}\right)^{3.4} \left(\frac{M_{\rm core}}{5M_\oplus}\right)^{1.7} \,.
\end{align}
This scaling relation is compared
against the numerical model in Figures
\ref{fig:gcr_t} and \ref{fig:gcr_M};
the agreement is good.

Without the fudge factor $f$ to mop up discrepancies, our derivation
states that conditions at the rcb are independent of
the nebular environment --- i.e., $T_{\rm rcb} = 2500$ K and
$\kappa_{\rm rcb} = \kappa ({\rm H}^-)$ are determined
by the microphysics governing the conversion of H$_2$ to H$^-$, and
$\rho_{\rm rcb}$ as given by equation (\ref{eq:rhorcb})
does not depend on the outer boundary conditions.
These statements are largely but not completely true.
In reality, there is a slight dependence of $\rho_{\rm rcb}$
on the nebular density $\rho_{\rm out}$ (as noted in paper I).
For the same core mass, GCR, and outer boundary radius,
a larger $\rho_{\rm out}$ implies a shallower atmospheric
density profile. Then the density at the rcb (whose
temperature is assumed fixed at $2500$ K)
should be lower;
in turn, the lower
$\rho_{\rm rcb}$ reduces the optical depth 
and thereby enhances the cooling luminosity
(equation \ref{eq:L}).
This explains qualitatively why the GCR (equivalently, $f$)
is a few times larger for the gas-rich case
than for the gas-poor case, all other
factors being equal.

Figure \ref{fig:gcr_t} also illustrates
how the threat of runaway gas accretion
is greater in gas-rich disks, not
only because they produce slightly faster cooling =
slightly faster accreting atmospheres, but also
because they last longer than gas-poor
disks. A 10$\times$ longer lifetime
enables the GCR to grow by an extra
factor of $10^{0.4} = 2.5$.
The final GCR shown for the gas-rich case
skirts dangerously close to the
runaway value
(formally evaluated to be 0.48; paper I). 
Although the curves for dusty atmospheres shown in
Figure \ref{fig:gcr_t} refer only to 5$M_\oplus$
cores at 0.1 AU, the same propensity to runaway applies
to dusty 5$M_\oplus$ planets at all orbital
distances out to 5 AU ($f$ hardly varies
between 0.1 AU and 5 AU).

Note that \citet{piso15} quoted a much larger
critical core mass of $\sim$30$M_\oplus$
at 5 AU in a dusty nebula
(see their Figure 7).
We have traced the origin of the 
discrepancy to three sources.
First, \citet{piso15} used the analytic
opacity model of \citet{bell94}
which, unlike the \citet{ferguson05} opacities
that we use, does not account for
different sublimation temperatures
of different dust species and appears
to overestimate $\kappa({\rm H}^-)$ at the rcb.
Their higher $\kappa$ suppresses
cooling relative to our models.
Second, these authors defined the runaway
time $t_{\rm run}$
as the moment when $M_{\rm gas}/\dot{M}_{\rm gas}$
falls to 10\% of its maximum value. This
time appears systematically longer than our
$t_{\rm run}$---defined as the time
when $L_{\rm cool}$ attains its
minimum---by factors of 2--3. 
We prefer our
definition as the minimum $L_{\rm cool}$
has physical significance: it
divides stable from unstable thermal 
equilibria in the presence of planetesimal accretion
(see discussion surrounding our Figure 1).
Finally, \citet{piso15} compared their
$t_{\rm run}$ against a disk lifetime
of $t_{\rm disk} = 3$ Myr, whereas
we adopt 
$t_{\rm disk} = 10$ Myr
for our gas-rich models.

\subsection{Dust-Free Atmospheres}
\label{ssec:dust-free-atm}

Dust-free atmospheres behave qualitatively
differently from dusty atmospheres.
Removing dust as a source of opacity (either
through grain growth or sedimentation; e.g.,
\citealt{mordasini14}; \citealt{ormel14}) renders
the outermost atmospheric layers entirely radiative.
The only rcb of the atmosphere
sits at the base of this radiative and nearly
isothermal outer shell:
$T_{\rm rcb} \sim T_{\rm out}$,
the temperature at the atmosphere's
outer boundary, set by the ambient disk.
Not surprisingly, how the GCR evolves
depends more
sensitively on nebular conditions for dust-free
atmospheres than for dusty atmospheres,
as the latter are buffered by the radiative
window opened by dust sublimation (Section 2.1).
Here we quote fitting formulae for $\kappa_{\rm rcb}$
and evaluate our GCR scaling relation (\ref{eq:gcr})
for dust-free atmospheres under various
nebular conditions.

\begin{figure}[!tbh]
\centering
\includegraphics[width=0.4\textwidth]{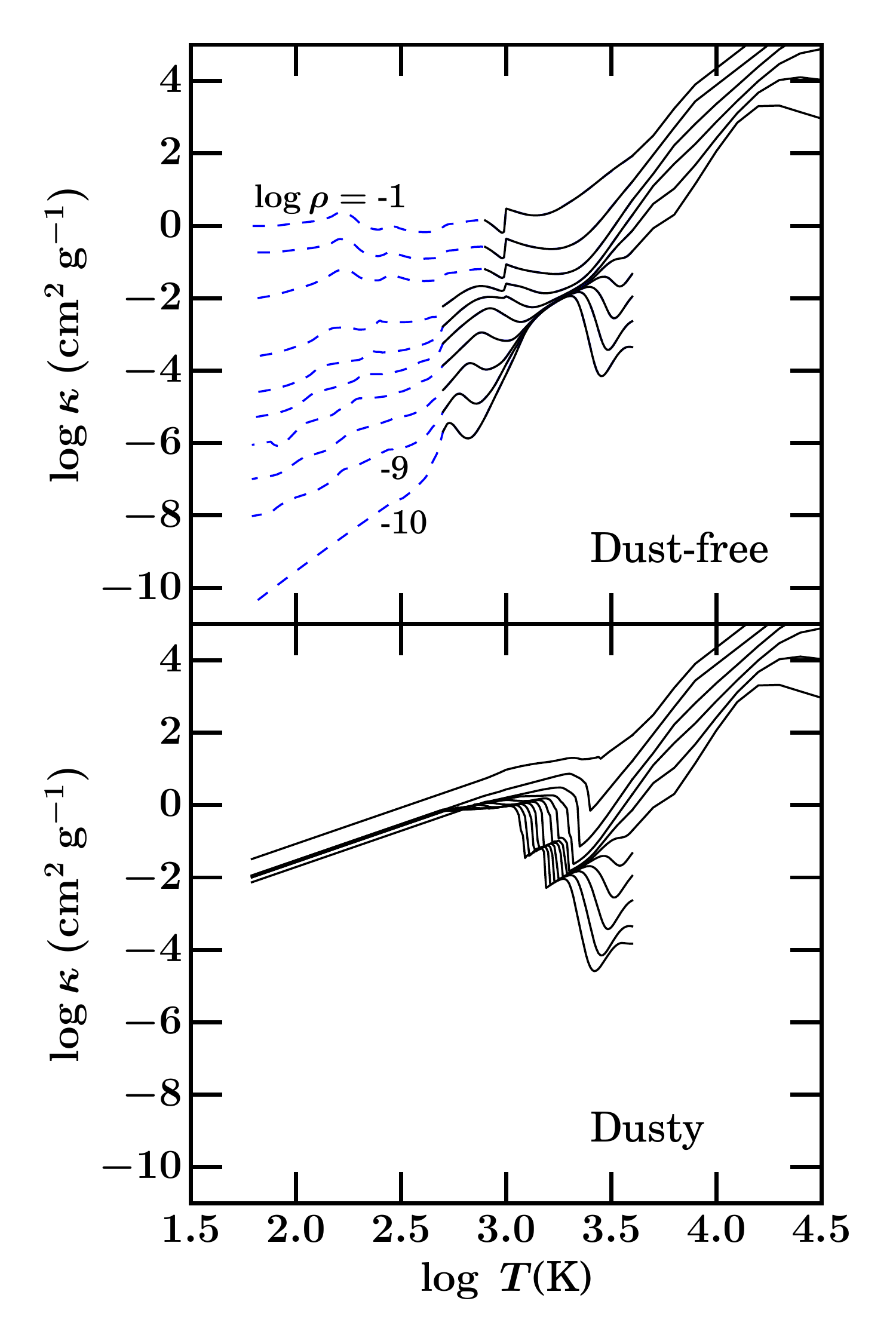}
\caption{\label{fig:opacity_table}
Dust-free (top) 
and dusty (bottom) opacities at solar metallicity for 
several densities ($\log\rho \,(\rhocgs) = -1$, -2,
..., -10 from top to bottom).
Black solid lines correspond to tabulated opacities
from \citet{ferguson05},
except for $\log\rho \geq -6$ and 
$\log \,T ({\rm K}) \geq 3.6$ where data are
extrapolated following the method described in
paper I. Blue dashed lines correspond to tabulated 
opacities from \citet{freedman14}, except 
for $\log\rho = -10$ and $\log T \leq 2.5$ 
where their analytic fits are used (their equations
3--5). \citet{freedman14}
present their opacity tables in ($P,T$) space so we
use $\mu=2.374$ to convert $P$ to $\rho$.
The abrupt jumps in the dust-free
$\kappa$ at $T = 1000$ K 
for $\log\rho \gtrsim -3$ are due to a failure
in equation-of-state calculations
(J.~Ferguson, private communication). 
Our model atmospheres never attain such high densities
at $T = 1000$ K so this bug is irrelevant.}
\end{figure}

The opacities underlying our models are
drawn from \citet{ferguson05}
and \citet{freedman14}.
We employ the former for
$\log \,T ({\rm K}) > 2.7$ (the domain of their
tabulations), and the latter for colder
temperatures. A smooth merging
of the two datasets 
is effected by offsetting all
of the data from \citet{freedman14}
to match the \citet{ferguson05} data
at $\log\, T = 2.7$. Depending on $\rho$,
the offset in
$\kappa$
ranges from 0.1 to 0.4 dex.
If no numerically tabulated value for $\kappa$
is available for a desired $(\rho,T)$ 
below $\log\, T = 2.7$,
we use the analytic fits given by
equations (3)--(5) of \citet{freedman14}; 
above $\log\, T = 3.6$,
we extrapolate
following the procedure described in paper I.

The top panel of Figure \ref{fig:opacity_table} 
plots the dust-free $\kappa(\rho,T)$.
We surmise the following
features. 
At $\log \,T \lesssim 3$, the main absorbers
are water, methane, and ammonia whose lines
are pressure broadened.
At $\log \,T \sim 3$, atomic alkali metals 
dominate the opacity \citep[see][their Figure 1]{freedman14}. The spike
in $\kappa$ that the alkalis produce is washed
out at large $\rho$, presumably from
pressure broadening. At still higher 
$\log \,T \gtrsim 3.3$, H$^-$ reigns, as it does for
dusty atmospheres.

Two-dimensional power laws fitted
to $\kappa (\rho, T)$ in restricted domains
of $(\rho, T)$ are presented below.
It is assumed throughout that $\kappa \propto Z$;
\citet{freedman14} found 
a nearly linear $Z$-dependence 
for $T \sim 250$--3000 K.

\subsubsection{Dust-Free and Gas-Poor at 0.1 AU}
\label{sssec:gcr-nog-dep-closein}
For dust-free atmospheres at 0.1 AU,
\begin{align}
\kappa (\operatorname{dust-free}, 0.1\, {\rm AU})  \simeq 7\times 10^{-3}~{\rm cm}^2~{\rm g}^{-1} \nonumber \\
\times 
\left(\frac{\rho}{10^{-5}\,\rhocgs}\right)^{0.3} 
 \left(\frac{T}{1000 \,{\rm K}}\right)^{0.9}\left(\frac{Z}{0.02}\right)^1
\end{align}
valid for $-5 \leq \log \,\rho\, (\rhocgs) \leq -3$ and 
$1000 \leq T ({\rm K}) \leq 2500$.
The corresponding GCR is 
\begin{align}
\label{eq:gcr-df-0.1au}
{\rm GCR} &\simeq 0.16 \left( \frac{f}{1.3} \right)
\left(\frac{t}{1\,{\rm Myr}}\right)^{0.4} \left(\frac{1600 \,{\rm K}}{T_{\rm rcb}}\right)^{1.9} \nonumber \\
&\,\,\,\,\,\,\,\times \left(\frac{0.02}{Z}\right)^{0.4} \left(\frac{\nabla_{\rm ad}}{0.17}\right)^{3.3}\left(\frac{\mu_{\rm rcb}}{2.37}\right)^{3.3}\left(\frac{M_{\rm core}}{5M_\oplus}\right)^{1.6}
\end{align}
where the nominal values for
$\nabla_{\rm ad}$ and $T_{\rm rcb}$ are
drawn from our full numerical model at 0.1 AU (for
comparison, $T_{\rm out} = 1000$ K),
and where we have calibrated $f = 1.3$.
Equation (\ref{eq:gcr-df-0.1au}) is plotted
against the full numerical solution in Figure
\ref{fig:gcr_t}.

The $\kappa$ quoted above is most relevant for
gas-poor disks at $\sim$0.1 AU whose surface
densities $\Sigma = \Sigma_{\rm MMEN}/200$.
The opacity behaves differently in gas-rich disks. 
We do not present results for dust-free atmospheres
in gas-rich disks at 0.1 AU because such
atmospheres stay fully convective for
GCRs at least up to $\sim$0.3
and cannot be evolved
using our numerical model. There would
not be much point to following them anyway,
since the combination of dust-free and
gas-rich conditions leads to rapid runaway.

\subsubsection{Dust-Free and Gas-Rich Beyond 1 AU}
\label{sssec:gcr-nog-full-1au}
Outside 1 AU, temperatures fall. 
For $100 \leq T({\rm K}) \leq 800$
and $-6 \leq \log \,\rho (\rhocgs) \leq -4$,
\begin{align}
\kappa (\operatorname{dust-free}, > 1 \,{\rm AU})  \simeq 1\times 10^{-5} \,{\rm cm^2 \,g^{-1}} \nonumber \\
\times \left(\frac{\rho}{10^{-6}\, \rhocgs}\right)^{0.6} \left(\frac{T}{100\,{\rm K}}\right)^{2.2}\left(\frac{Z}{0.02}\right)^1
\end{align}
whence
\begin{align}
\label{eq:gcr-df-1au}
{\rm GCR} &\sim 0.1 \left( \frac{f}{2.8} \right) \left(\frac{t}{1\,{\rm kyr}}\right)^{0.4}\left(\frac{200\,{\rm K}}{T_{\rm rcb}}\right)^{1.5} \nonumber \\
&\times \left(\frac{0.02}{Z}\right)^{0.4} \left(\frac{\nabla_{\rm ad}}{0.25}\right)^{2.2}\left(\frac{\mu_{\rm rcb}}{2.37}\right)^{2.2}\left(\frac{M_{\rm core}}{5M_\oplus}\right)^{1}
\end{align}
where the nominal $\nabla_{\rm ad} = 0.25$
(equivalently, $\gamma = 1.3$) represents
a rough average over our dust-free numerical
models at distances $> 1$ AU.
The numerical models also indicate that at these large
distances, $T_{\rm rcb}$ nearly equals 
$T_{\rm out}$; more distant planets have
more nearly isothermal outer layers.
Equation (\ref{eq:gcr-df-1au}),
normalized for nebular conditions at 5 AU,
is tested against the full numerical solution
in Figure \ref{fig:gcr_t}.

Equation (\ref{eq:gcr-df-1au}) highlights
just how susceptible dust-free super-Earths
are to runaway at large distances in gas-rich
disks. At 5 AU,
$T_{\rm rcb} = T_{\rm out} \simeq 200$ K
and a $5 M_\oplus$ core will attain GCR $\sim$ 0.5
in as short a time as 0.05 Myr.
(We do not present solutions for gas-poor disks
at large distances because the relevant gas densities
fall below those covered by our opacity tables.
It is easy, however, to guess what these solutions
would look like: GCRs would increase
on timescales almost as fast as those in gas-rich
disks, and would be limited only by the
gas available and the remaining disk lifetime.)

\section{Summary}
\label{sec:conclusion}

We have developed an analytic model 
describing how a solid core accretes
gas from its parent disk. 
The model
assumes that the planet's growing
atmosphere cools passively: 
heat inputs from external sources
such as planetesimal accretion are ignored.
Heating from planetesimal accretion
can be relevant for low-mass $< 5 M_\oplus$
cores, but only under exceptionally fine-tuned
conditions. Too high a planetesimal
accretion rate leads to excessively massive
cores which undergo runaway, while too low an accretion rate
is energetically irrelevant. 
The window for planetesimal accretion to be relevant is so narrow
--- it vanishes completely for core masses $> 5 M_\oplus$,
which cannot help but undergo runaway in gas-rich disks
that persist for $\sim$10 Myr ---
that neglecting energy deposition by solids is almost certainly
safe.

We derived how
the gas-to-core mass ratio (GCR) 
varies with time $t$, 
core mass $M_{\rm core}$, and metallicity $Z$
for passively cooling atmospheres.
The scaling relationship, 
given by equation (\ref{eq:gcr}), 
applies in various formation environments
at times preceding the onset of runaway
gas accretion (at times when
the envelope's self-gravity is still
negligible, i.e., when GCR $< 0.5$).
The analytic results agree with 
numerical calculations to within 
factors of 1--3.

Three physical ingredients 
make our derivation possible.
(1) To cool is to accrete:
in the absence of any heat source, 
atmospheres of mass $M_{\rm gas}$
grow on a timescale equal 
to their cooling time:
$M_{\rm gas}/\dot{M}_{\rm gas} \sim |E|/L_{\rm cool} \sim t$,
where $E$ is the atmosphere's energy
and $L_{\rm cool}$ is its cooling luminosity. 
The final GCR is then set by the condition 
that the atmosphere's cooling time $t_{\rm cool}$
equals the gas disk's depletion time $t_{\rm disk}$.
(2) Most of the mass of the atmosphere
is concentrated toward the core,
whose given properties 
provide an easy reference for scaling
our variables. The mass concentration follows
from density gradients made steep
by the dissociation of H$_2$ and
the consequent lowering of the adiabatic
index $\gamma$ down to 1.2--1.3. 
(3) The cooling luminosity is regulated by the 
radiative-convective boundary (rcb) and therefore 
can be evaluated there.
Specifying the microphysical
properties of this boundary (temperature, opacity)
enables us to quantify the rate of cooling.

The scaling indices of
GCR ($t$, $M_{\rm core}$, $Z$)
are determined by the innermost convective
zone's adiabatic index
$\gamma$ (calibrated using our numerical
models) and the
dependencies of the opacity at the rcb.
Dusty atmospheres have rcb's that are distinct
from those of
atmospheres rendered dust-free by grain
growth and sedimentation.
In dusty atmospheres (Section
\ref{ssec:dusty-atm}), the
rcb occurs universally --- at all orbital
distances and in gas-rich or gas-poor nebulae ---
at the H$_2$ dissociation
front, setting the temperature
$T_{\rm rcb} \simeq 2500$ K.
The rcb in dusty atmospheres is
insensitive to external nebular
conditions because 
dust sublimation opens a radiative window
in the interior of the planet, whose
base (the rcb) is determined by the local
microphysics of how H$_2$ converts into
H$^-$ whose opacity rises sharply with temperature.
All dusty atmospheres evolve as GCR
$\propto t^{0.4} M_{\rm core}^{1.7} Z^{-0.4} \mu_{\rm rcb}^{3.4}$: 
the scaling indices 
apply globally.
Note that the mean molecular
weight $\mu$ depends inversely on $(1-Z)$
so that all other factors being equal,
GCR first decreases with $Z$
(as $Z^{-0.4}$; this dependence 
arises from opacity) and then increases
(via $\mu^{3.4}$) as atmospheres become ``heavier".

Dust-free atmospheres (Section
\ref{ssec:dust-free-atm}) are more transparent
in their outer layers and their
rcb's are located
at higher altitudes: conveniently,
$T_{\rm rcb} \sim T_{\rm out}$, the given
temperature of the ambient nebula.
For dust-free atmospheres,
the dependence of rcb properties
on nebular properties implies that
the GCR scaling indices 
change with orbital distance,
depending on the behavior
of the local opacity.
We examined two cases.
At 0.1 AU in a gas-poor nebula, 
GCR $\propto t^{0.4} T_{\rm rcb}^{-1.9} M_{\rm core}^{1.6} Z^{-0.4} \mu_{\rm rcb}^{3.3}$.
Beyond 1 AU in a gas-rich nebula, 
GCR $\propto t^{0.4} T_{\rm out}^{-1.5} M_{\rm core}^1 Z^{-0.4} \mu_{\rm rcb}^{2.2}$. 
For all the cases we evaluated,
GCR $\propto t^{0.4}$. 
The GCR-$t$ scaling is determined by
the $\kappa_{\rm rcb}$-$\rho_{\rm rcb}$ 
scaling (see equation \ref{eq:gcr}).
It appears the latter scaling does not change
much across environments ($\kappa_{\rm rcb}
\propto \rho_{\rm rcb}^{\alpha}$ where
$\alpha \simeq 0.3$--0.6;
this appears to be an average
between $\alpha = 1$ characterizing
pressure-broadened opacities at $T \sim 100$ K
and $\alpha \approx 0$ characterizing
the opacity at $T \sim 2000$ K;
see Figure 7 of \citealt{freedman14}).

While dusty atmospheres behave
more-or-less the same way under a variety
of nebular conditions, dust-free atmospheres
depend more sensitively on disk
temperatures. Gas opacities tend
to decrease with colder temperatures and
therefore dust-free atmospheres grow
faster the farther they are from their central
stars.

Atmospheres accrete faster
in gas-rich environments than in gas-poor
ones, but not by much.
For example, in our numerical models
of dusty atmospheres,
dropping the nebular density by a factor
of 200 drops the GCR by a factor of 2.5,
all other factors being equal. We have checked
that similar results obtain for dust-free
atmospheres.
Without introducing a normalization
correction (our factor of $f$
in equation \ref{eq:gcr}),
our analytic scalings for dusty atmospheres would
predict no dependence at all
on the ambient nebular density (see
Section \ref{ssec:gcr-gas-rich}
for the technical reason why
in reality there is a small dependence).
The insensitivity to nebular density
suggests that the accretion of planetary 
atmospheres proceeds about the same
way whether or not planets open gaps in disks.
Gap opening in viscous gas disks 
does not starve the planet; it is
well-recognized that material
continues to flow past the planet, through
the gap~\citep[e.g.,][]{lubow06,duffell14}.
Moreover, the factors by which densities
are suppressed in gaps are likely overestimated
in 2D simulations \citep[e.g.,][]{kanagawa15}:
3D gaps are messier.
Careful studies of the 
3D flow dynamics of planets embedded
in disks \citep[e.g.,][]{d'angelo13, ormel15, fung15} 
are clearly needed to test whether
gap opening and/or hydrodynamic effects substantively
alter the 1D accretion theory we have presented.

The analytic solutions for gas-to-core ratios
developed here enable us to calculate the
properties of planetary envelopes just before
protoplanetary disk gas disperses.
The solutions --- encapsulated
in the condition $t_{\rm cool} \sim t_{\rm disk}$ ---
provide initial conditions
for models of subsequent atmospheric loss
by hydrodynamic winds powered by stellar
irradiation or internal heat
\citep{owen13,owen15}.
Our model complements studies 
that include detailed molecular chemistry
\citep[e.g.,][]{hori11,venturini15};
we find good agreement with those models
for $Z$ up to $\sim$0.5.
The simplicity of our solutions make them
suitable for the construction and/or
diagnosis of more realistic
models that include 3D hydrodynamic effects
\citep[e.g.,][]{d'angelo13,ormel15,fung15}.
The solutions can also be used to quickly assess
the plausibility of various planet
formation scenarios. This is the task we set
ourselves for paper III.
\vspace{0.2in}

\acknowledgments
Our high-$Z$ analysis was made possible by the
generous contribution of Jason Ferguson who
calculated all the opacity tables for 
$\log \,T \, ({\rm K}) > 2.7$. 
We are grateful to Brad Hansen,
Chris Ormel, James Owen, and Andrew Youdin
for thoughtful and encouraging comments on our manuscript,
and Gennaro D'Angelo,
Rebekah Dawson, Jonathan Fortney, Michael Line, 
Eric Lopez, Mark Marley, Ruth Murray-Clay, Julia Venturini, and Yanqin Wu 
for helpful and motivating discussions.
We thank the referee for providing an especially inspiriting
report that led to improvements in the manuscript.
EJL is supported in
part by the Natural Sciences and 
Engineering Research Council of Canada under PGS D3
and the Berkeley Fellowship. EC acknowledges
support from grants AST-0909210 and AST-1411954
awarded by the National Science Foundation, NASA
Origins grant NNX13AI57G, and Hubble Space
Telescope grant HST-AR-12823.001-A.
Numerical calculations were performed on the SAVIO 
computational cluster resource provided by the 
Berkeley Research Computing program at the 
University of California Berkeley, supported by 
the UC Chancellor, the UC Berkeley Vice Chancellor 
for Research, and Berkeley's Chief Information
Officer.

\bibliography{gcr}

\end{document}